\documentclass[onecolumn]{IEEEtran}
\usepackage{cite}
\usepackage{verbatim}
\usepackage{multirow}
\usepackage{graphicx}
\usepackage[cmex10]{amsmath}
\usepackage{stfloats}
\usepackage{minipage-marginpar}
\usepackage{subfig}
\usepackage[none]{hyphenat}
\begin{document}
\title{Growth Patterns of US Children from 1963 to 2012}
\author{\IEEEauthorblockN{Xiang Zhong\IEEEauthorrefmark{1},
Jingshan Li\IEEEauthorrefmark{2},
Goutham Rao\IEEEauthorrefmark{3}, and
KP Unnikrishnan\IEEEauthorrefmark{4}}\\
\IEEEauthorblockA{\IEEEauthorrefmark{1}
University of Wisconsin-Madison,
Madison, Wisconsin 53706--1539\\ Email: xzhong4@wisc.edu}\\
\IEEEauthorblockA{\IEEEauthorrefmark{2}
University of Wisconsin-Madison,
Madison, Wisconsin 53706--1539\\ Email: jingshan@engr.wisc.edu}\\
\IEEEauthorblockA{\IEEEauthorrefmark{3} NorthShore University HealthSystem,
Evanston, Illinois 60201-3137\\ Email: grao@northshore.org}\\
\IEEEauthorblockA{\IEEEauthorrefmark{4}NorthShore University HealthSystem,
Evanston, Illinois 60201-3137\\ Email: kunnikrishnan@northshore.org}}

\maketitle
\begin{abstract} Anthropometric measurements such as weight, stature (height), and body mass index (BMI) provide reliable indicators of children's growth. The 2000 CDC growth charts are the national standards in the United States for these important measures. But these growth charts were generated using data  from 1963-1994. To understand the growth patterns of US children since 1994, we generate weight-for-age, stature-for-age and BMI-for-age percentile curves for both boys and girls aged 2-20 through the methods used to generate the 2000 CDC growth charts. Our datasets are from the National Health and Nutrition Examination Survey (NHANES) for years 1999-2010 and and from NorthShore University HealthSystem's Enterprise Data Warehouse (NS-EDW) for years 2006-2012. The weight and BMI percentile curves generated from NS-EDW and NHANES data differ substantially from the CDC percentile curves, while those for stature do not differ substantially. We conclude that the population weight and BMI values of US children in recent years have increased significantly since 2000 and the 2000 CDC growth charts may no longer be applicable to the current population of US children. Our charts poignantly reveals the increasing obesity of American children.
\end{abstract}

\vspace*{0.1in}


\section{Introduction}

Childhood obesity has become a national epidemic in the United States. An estimated 31.7\% of all American children ages 2-19 are currently either overweight or obese \cite{Ogden10}. Childhood obesity is associated with serious medical, psychological and social consequences in adult life.

Anthropometric data are valuable objective indicators of physical growth in children \cite{CDCpaper02}. Growth charts generated from it consist of a series of percentile curves that illustrate the distribution of selected body measurements including weight, stature and the body mass index (BMI) in children. Pediatric growth charts have been used by pediatricians, nurses, and parents to track the growth of infants, children, and adolescents in the United States since 1977 \cite{CDCwebsite}. A variety of growth reference both data based and method oriented were developed and published (see representative studies \cite{Owen78, WHO, Cole98, Cole99, Cole88, Cole92, Flegal13, ChenSAS}). Body mass index (BMI) is a general measure calculated from person's weight and height. It is defined as the ratio of weight ($kg$) to squared height ($m^2$) and is popularly used as a measure of overweight and obesity \cite{Mei02}. For children and teenagers, sex- specific BMI follows a characteristic pattern. It increases rapidly during the first year of life and then declines, reaching a nadir and increasing again \cite{Rolland84}. This pattern is observed (with minor differences) in both boys and girls. The inflection point at which BMI begins to increase again is known as the adiposity rebound (AR) \cite{Rolland87}. Adiposity rebound normally occurs at roughly six years of age. A significant body of evidence
indicates that early adiposity rebound (i.e., before age six) is a significant predictor of overweight and obesity in adolescence and young adulthood \cite{Taylor05}.

The percentiles of BMI for a specified age is of particular interest in light of public health concerns. Based on the CDC growth charts, a BMI percentile $\geq 85$ and $<95$ is classified as overweight and a BMI percentile $\geq 95$ is is classified as obese; the lower percentiles are observed for issues associated with being underweight \cite{Flegal09,Tim07}. However, these standards used to identify obesity have serious shortcomings including lack of accuracy associated with increased health risk \cite{Rao13}. The CDC has not published updated growth charts since 2000, while more recent data has become available from NHANES and in systems with electronic health records like NorthShore Enterprise Data Warehouse (NS-EDW).

In this paper, our objective is to determine if we could produce growth charts based on a large pool of recently collected electronic data (NorthShore Enterprise Data Warehouse (NS-EDW) data from 2006 to 2012  and National Health and Nutrition Examination Survey (NHANES) data from 1999 to 2010), and to investigate the newer growth curves' variation from the established CDC curves in terms of shape or patterns of growth. By closely following the methods used to generate the 2000 CDC growth charts, growth charts for these two data sets are generated and compared with the CDC charts. We conclude that at the same age of sampling, the population weight and BMI values increased by year of sampling, which indicates an increasing obesity of American children, and the CDC growth charts may no longer be accurate standards for today's population of American children.

The remainder of this paper is structured as follows. The datasets and methods used to generate the growth charts are described in Section 2. Section 3 summarized the results. Discussions are presented in Section 4 and conclusions are formulated in Section 5, respectively.
\section{Materials and Methods}
In this paper, the growth charts were developed using data from the NorthShore Enterprise Data Warehouse (NS-EDW) and National Health and Nutrition Examination Survey (NHANES). Closely following the methods used to generate the 2000 CDC growth charts, selected empirical percentiles were generated and smoothed through a variety of parametric and non-parametric procedures. For each data set, weight-for-age, stature-for-age and BMI-for-age percentiles for boys and girls of age 2-20 years were generated separately and compared with the 2000 CDC growth curves. Further more, to illustrate the variation trend over years of sampling, the NHANES data were separated into three non-overlapping four-yearly data sets. Five boys weight-for-age growth charts generated from the 2000 CDC, NHANES 99-02, NHANES 03-06, NHANES 07-10, and NS-EDW 06-12 data are compared.

\subsection{Data sources}
\begin{figure}[!t]
\centering
\includegraphics[width=3in]{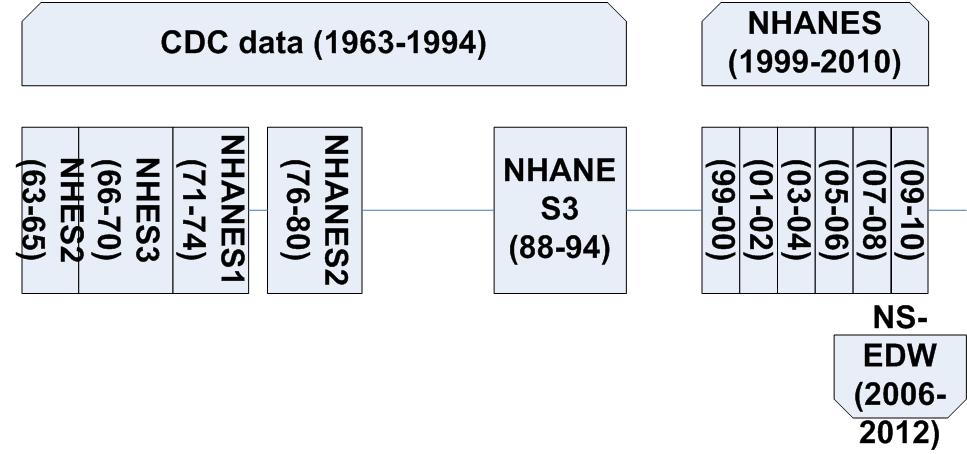}
\caption{Time line for data sources}
\label{fig_timeline}
\end{figure}

The general information of each data source and the corresponding charts they generated are summarized in Table~\ref{table_figlist}. The time line of all the data source (CDC data, NS-EDW and NHANES data) is illustrated in Figure~\ref{fig_timeline}. The information stratified by sex, and race/ethnicity for the NS-EDW and NHANES data set are described in detail separately. A detailed description of the CDC data set is given in  \cite{CDCpaper02}.

\begin{table}[!t]
\renewcommand{\arraystretch}{1.3}
\caption{Data characteristics}
\label{table_figlist}
\centering
\begin{tabular}{|c||c||c||c||c|}
\hline
Data Set & Year & Subject & Sex & Chart\\ \hline
NS-EDW Data & & & &  \\ \hline
NS-EDW & 2006-2012 &  Age:2-20 &M,F & W,S,BMI\\ \hline
NHANES Data & & & &  \\ \hline
NHANES & 1999-2010 &  Age:0-26 &M,F & W,S,BMI\\ \hline
NHANES & 1999-2002 &  Age:0-26 &M & W\\ \hline
NHANES & 2003-2006 &  Age:0-26 &M & W\\ \hline
NHANES & 2007-2010 &  Age:0-26 &M & W\\ \hline
CDC Data & & & &  \\ \hline
NHES2  & 1963-1965 &  Age:6-12 &M,F & W,S,BMI\\ \hline
NHES3  & 1966-1970 &  Age:12-18 &M,F & W,S,BMI\\ \hline
NHANES1  & 1971-1974 &  Age:1-20 &M,F & W\\ \hline
NHANES1  & 1971-1974 &  Age:2-25 &M,F & S,BMI\\ \hline
NHANES2  & 1976-1980 &  Age:1-20 &M,F & W\\ \hline
NHANES2  & 1976-1980 &  Age:2-25 &M,F & S,BMI\\ \hline
NHANES3  & 1988-1994 &  Age:1-6  &M,F & W\\ \hline
NHANES3  & 1988-1994 &  Age:2-25 &M,F & S\\ \hline
NHANES3  & 1988-1994 &  Age:2-6  &M,F & BMI\\ \hline
\end{tabular}
\end{table}

\subsection{NorthShore Enterprise Data Warehouse Data}
Children's weight and stature are routinely measured at annual well-child visits and at other visits to physicians. In state-of-the-art clinical informatics systems such as the NorthShore University HealthSysetem's (NS), an Enterprise Data Warehouse (EDW) is in place to capture clinical and administrative data for quality improvement and research. The foundation of this system is a comprehensive electronic medical record using the EPIC platform that has been in place since 2003 and extends across 4 hospitals accounting for over 1000 beds and 60,000 annual admissions; 4 emergency departments seeing over 100,000 annual visits; and 80 office practices with over one million encounters per year. All encounters use a common
database and are live on the system. A query of the NS-EDW was completed to quantify the number of children in whom weight and stature information are feasible with multiple BMI measurements over time. Data from more than 400,000 encounters, stretching from years 2006 through mid 2012 are collected and all data are available for research in fully de-identified form. 
\subsubsection{Data Exclusion}
Several exclusions were made prior to data processing. NS-EDW data missing weight, stature or BMI information are excluded. Weight larger than 200 $kg$ and stature larger than 242 $cm$ are excluded, which are outliers due to inaccurate measurements or recordings. BMI value less than 6 $kg/m^2$ or larger than 100 $kg/m^2$ are excluded. Besides, data with age under two are cut off the data set for missing accurate recordings of stature. In order to get representatively longitudinal data, children have encounters less than five times are filtered.
\subsubsection{Data Statistics}
The query results after data exclusion were approximately 7592 boys and 6878 girls age 2-20 years with a minimum of five BMI measurements separated in time. The BMI value ranges from 6.7 to 63.1 ($kg/m^2$). The detailed demographic information for the filtered NS-EDW data is summarized in Table~\ref{table_nsdata}.

\begin{table}[!t]
\renewcommand{\arraystretch}{1.3}
\caption{Demographic information for NS-EDW data}
\label{table_nsdata}
\centering
\begin{tabular}{|c||c||c|}
\hline
       & Boys  & Girls \\ \hline
Encounters &  50775 & 45390\\ \hline
Patients & 7592 & 6878 \\ \hline
Race/Ethnicity & & \\ \hline
African American &5.9\% & 5.5\% \\ \hline
American Indian & 0.6\% & 0.7\%\\ \hline
Asian & 3.6\% &4.5\% \\ \hline
Caucasian &57.8\% & 57.9\%\\ \hline
Hispanic/Latino &5.0\% & 4.8\% \\ \hline
Other &27.0\%& 26.6\%\\ \hline
Measurement&& \\ \hline
Wt(kg) &7.2-180.1& 4.2-163.2\\ \hline
St(cm) &60.9-241.3 & 53.0-221.0\\ \hline
BMI($kg/m^2$) &6.73-63.11 & 6.76-56.69\\ \hline
\end{tabular}
\end{table}

\subsection{National Health and Nutrition Examination Survey Data}
The National Health and Nutrition Examination Survey (NHANES) is a program of studies designed to assess the health and nutritional status of adults and children in the United States. The survey combines interviews including demographic, socioeconomic, dietary, and health-related questions  and physical examinations consist of medical, dental, and physiological measurements, as well as laboratory tests administered by highly trained medical personnel. NHANES findings are the basis for national standards for such measurements as height, weight, and blood pressure. Data from this survey are widely used in epidemiological studies and health sciences research, which help develop sound public health policy, direct and design health programs and services, and expand the health knowledge for the Nation \cite{CDCnhaneswebsite}.
\subsubsection{Data Exclusion}
Data from the six national surveys (09-00, 01-02, 03-04, 05-06, 07-08, 09-10) were pooled to construct growth charts. To achieve better precision of the empirical percentiles, pooling is introduced to enhance the number of subjects at each age, thereby increasing the stability of the outlying percentile estimates.

Similarly, the NHANES data were first filtered by encounters whose age are within the scope of study (0-26 years). Then, data missing weight, stature or BMI information were excluded. Besides, extremely high weight (larger than 300 $kg$) and BMI (larger 100 $kg/m^2$) are also excluded.
\subsubsection{Data Statistics}
The NHANES data has 62160 encounters in total with around 10,000 encounters for each bi-yearly data set. After the exclusion of missing value, there were 11820 boys and 11538 girls with complete information about age at sampling, weight, stature and BMI. The BMI value ranges from 7.99 to 66.32 ($kg/m^2$). The demographic information for
the filtered NHANES data in detail is shown in Table~\ref{table_nhdata}.

\begin{table}[!t]
\renewcommand{\arraystretch}{1.3}
\caption{Demographic information for NHANES data}
\label{table_nhdata}
\centering
\begin{tabular}{|c||c||c|}
\hline
       & Boys  & Girls \\ \hline
Encounters &  11820 & 11538\\ \hline
Ethnicity & & \\ \hline
Mexican American &31.2\% &31.8\% \\ \hline
Other Hispanic & 6.3\%&6.4\%\\ \hline
Non-hispanic White &28.5\%&27.9\% \\ \hline
Non-hispanic Black &29.2\% &28.5\% \\ \hline
Other multi-racial &4.8\% & 5.3\%\\ \hline
Measurement&& \\ \hline
Wt(kg) &9.7-239.4& 8.9-174.8\\ \hline
St(cm) &79.0-204.4 & 78.0-187.2\\ \hline
BMI($kg/m^2$) &11.98-66.32 & 7.99-62.08\\ \hline
\end{tabular}
\end{table}

\subsection{Statistical Curve Smoothing Procedures}
To generate the CDC-like boys and girls weight-for-age, stature-for-age and BMI-for-age percentiles with NS-EDW and NHANES data, we replicated the statistical procedures described in \cite{CDCpaper02} using custom written computer programs in R \cite{Rwebsite}. These programs were applied to generate empirically selected percentiles. Custom-written curve smoothing methods that closely follow those used by CDC \cite{CDCpaper02} were then applied to the major percentiles charts. The procedures for each stage are described in detail for each growth chart in the following subsections.
\subsubsection{Age Groupings and Curve Smoothing Stages}
The empirical percentiles with grouped age provide a discrete approximation for the population percentiles. To generate the initial selected major percentiles, data with age range from two to twenty were separated into 36 half-year age groups. Each age group was categorized by the midpoint of an age range. For example, age 2.25 years included ages from 2.0 years to 2.5 years of age. This pattern started from 2.25-year-age and continued to the 19.75-year-age interval. (Detailed description please refer to \cite{CDCpaper02}) The basic percentile definition $n=P/100\cdot N+1/2$ (where $N$ is the data size and $P$ is the desired percentile) and round $n$ to the nearest integer was used to get the corresponding percentile value.

For each growth chart, the initial smoothing methods were applied to nine empirical percentiles (3rd, 5th, 10th, 25th, 50th, 75th, 90th, 95th, and 97th). In addition, the 85th percentile was included in the BMI-for-age charts because the 85th percentile of BMI has been recommended as a cutoff to identify children and adolescents who are overweight \cite{Himes94, Barlow98}.

The empirical percentile points at the midpoint of each age group were calculated initially. The irregular plots of empirical percentile values were smoothed then to produce clinically useful percentile curves. Followed the methods in the CDC paper, several different approaches were used in the smoothing stage. Stature-for-age was smoothed using a nonlinear model. The empirical percentiles for weight and BMI were initially smoothed using locally weighted regression (LWR) \cite{Cleveland79} and then smoothed using a family of polynomial regression.
\subsubsection{Weight-for-age 2 to 20 years}
In order to compensate the insufficient sample size for age around and above 20 and below 2, some empirical data less than 2 years of age and greater than or equal to 20 years were included to make the curves smooth and consistent. This compound data set was initially smoothed with a locally weighted regression (LWR) procedure \cite{Cleveland79}. The LWR provides an intermediate smoothed curve for further parametric smoothing. LWR applies a weight function to data in the neighborhood of the value to be estimated. Ages at measurements that are near that of the value to be estimated received larger weights than those farther away from the specific age \cite{CDCpaper02}. The width of the LWR moving window was chosen to balance the degree of smoothness and fidelity to the data. Based on the shape of the underlying empirical percentile curves, the smoothed value was estimated by a LWR on 10-15 neighborhood points adjacent accordingly.

For NS-EDW boys and girls, the weight value at year 2 were repeated seven times from age 1.75 down to 1.15 years by 0.1-year interval. When smoothing the value at age 20 years, seven data points at ages over 20 years were necessary for moving window with width of 15 neighborhood points. The maximum value through age 16.75 to 19.75 were repeated seven times from 20.25 to 23.25 years by 0.5-year interval. For NHANES boys and girls, the weight value at year 1.75 were repeated seven times from age 1.75 down to 1.15 years by 0.1-year interval. The data for age above 20 is available for NHANES data so the empirical percentiles from 20.25 to 23.25 years by 0.5-year intervals were used as seven additional data points for smoothing. Same methods were applied to the three non-overlapping NHANES data sets. After the LWR procedure, the percentile curves were fit by a 10-degree polynomial for both boys and girls. The parameters of the polynomial regression for nine percentiles are estimated.
\subsubsection{Stature-for-age 2 to 20 years}
The stature-for-age curves for ages from 2 to 20 years were smoothed with a nonlinear model following the format shown in CDC paper:
\begin{equation*}
f(t)=\frac{a_1}{1+e^{-b_1(t-c_1)}}+\frac{a_2}{1+e^{-b_2(t-c_2)}}+\frac{a_3}{1+e^{-b_3(t-c_3)}}
\end{equation*}
where $f(t)$ is stature in cm, $t$ is age in years (calculated as midpoint of age range) and $a_1$, $b_1$, $c_1$, $a_2$, $b_2$, $c_2$, $a_3$, $b_3$, $c_3$, are specific for each of nine percentiles being smoothed. This model ensured a monotonic increase in stature. However, some of the empirical percentile curves derived from both data sets were irregular due to sampling variations and small number of subjects of age 18 to 20. To aid in smoothing the irregular empirical percentile curves, especially the 95th and 97th percentiles, weighted linear regression were applied for those non-smoothing curves before being fit into parametric models.
\subsubsection{ BMI-for-age 2 to 20 years}
Ten empirical percentiles were calculated for the BMI-for-age charts. The additional 85th percentile was required for boys and girls to identify children and adolescents overweight. Each smoothed value was estimated by weighted linear regression on the five neighborhood points adjacent to the value to be estimated from ages 2 to 12.5 years. At the lower end, two additional points were needed in the smoothing window. A neighborhood point of age 2 years was used, this value was also used for age 1.75 year. At the upper end, for the NS-EDW data, from ages 13 to 20, a 20-points smoothing procedure was applied. The maximum BMI values in each empirical percentile from 16.75 to 19.75 years were chosen and repeated in 0.5-year interval from 20.25 through 23.25. For the NHANES data, the maximum BMI values in each empirical percentile from 19.75 through 25.25 years were chosen and repeated in 0.5-year interval from 20.25 through 25.75 years. As clarified in the CDC paper, taking maximum values as additional data in smoothing the windows ensured that the BMI curves did not increase beyond the maxima at the upper ends of the age ranges. The smoothed percentile curves obtained through LWR were then fit by a four-degree polynomial regression to achieve parametric percentiles.
\section{Results}
\begin{figure}[!t]
\centering
\includegraphics[width=3.5in]{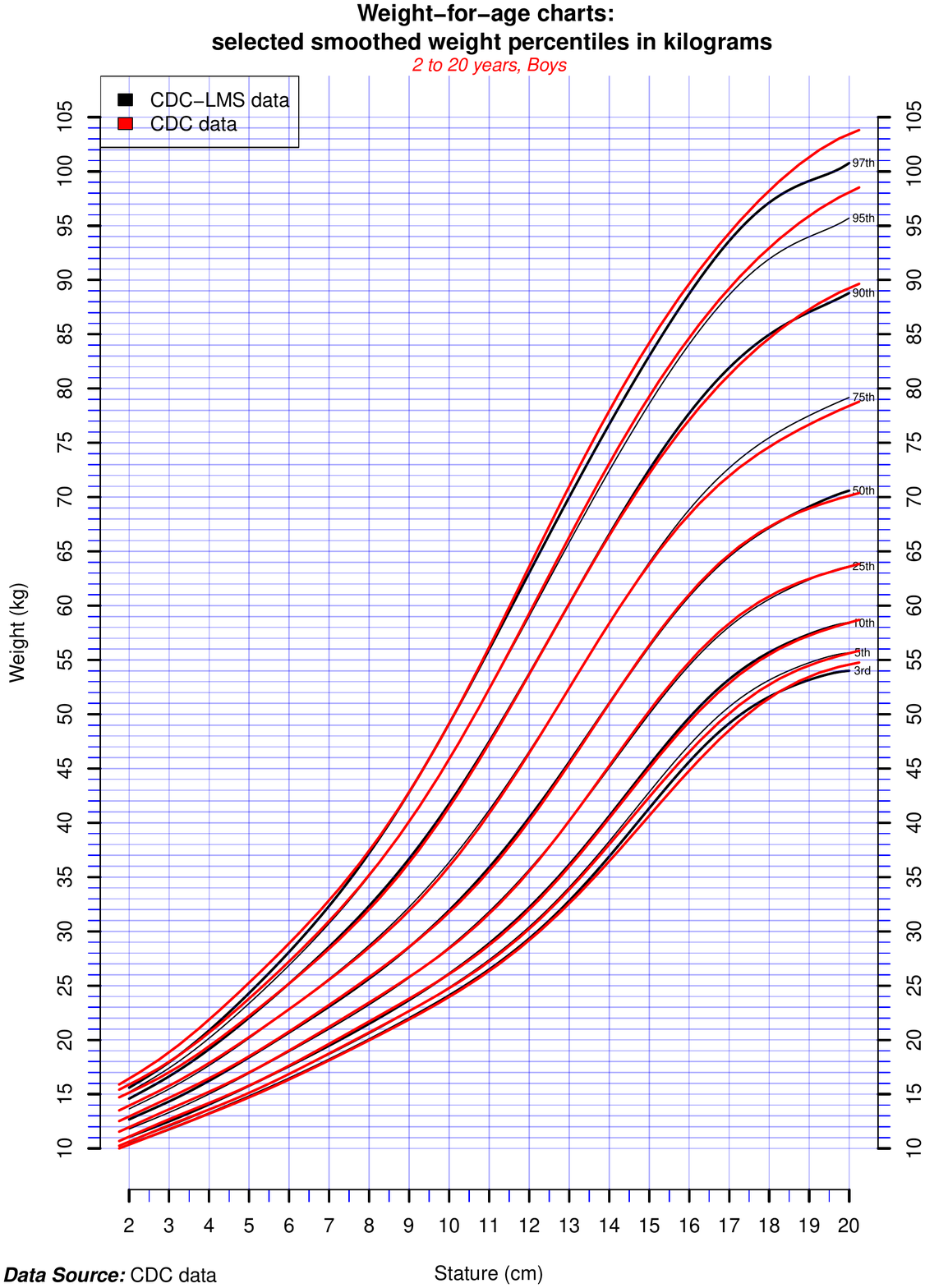}
\caption{CDC boys weight-for-age percentiles using two methods}
\label{fig_cdc}
\end{figure}

To address the success in reproducing the CDC curve smoothing procedures, the boys weight-for-age percentile curves generated using part of the CDC data source and the curves generated by the published LMS value from the CDC website (see \cite{CDCdatawebsite}) are compared and shown in Figure~\ref{fig_cdc}.

The result showed that the percentiles generated by closely following the CDC methods using partly the CDC data source is very similar to the published CDC percentiles. The slight difference might due to incomplete data sets and insufficient outlier exclusion.

\begin{figure*}[p]
\centerline{\subfloat[NS-EDW]{\includegraphics[width=3.5in]{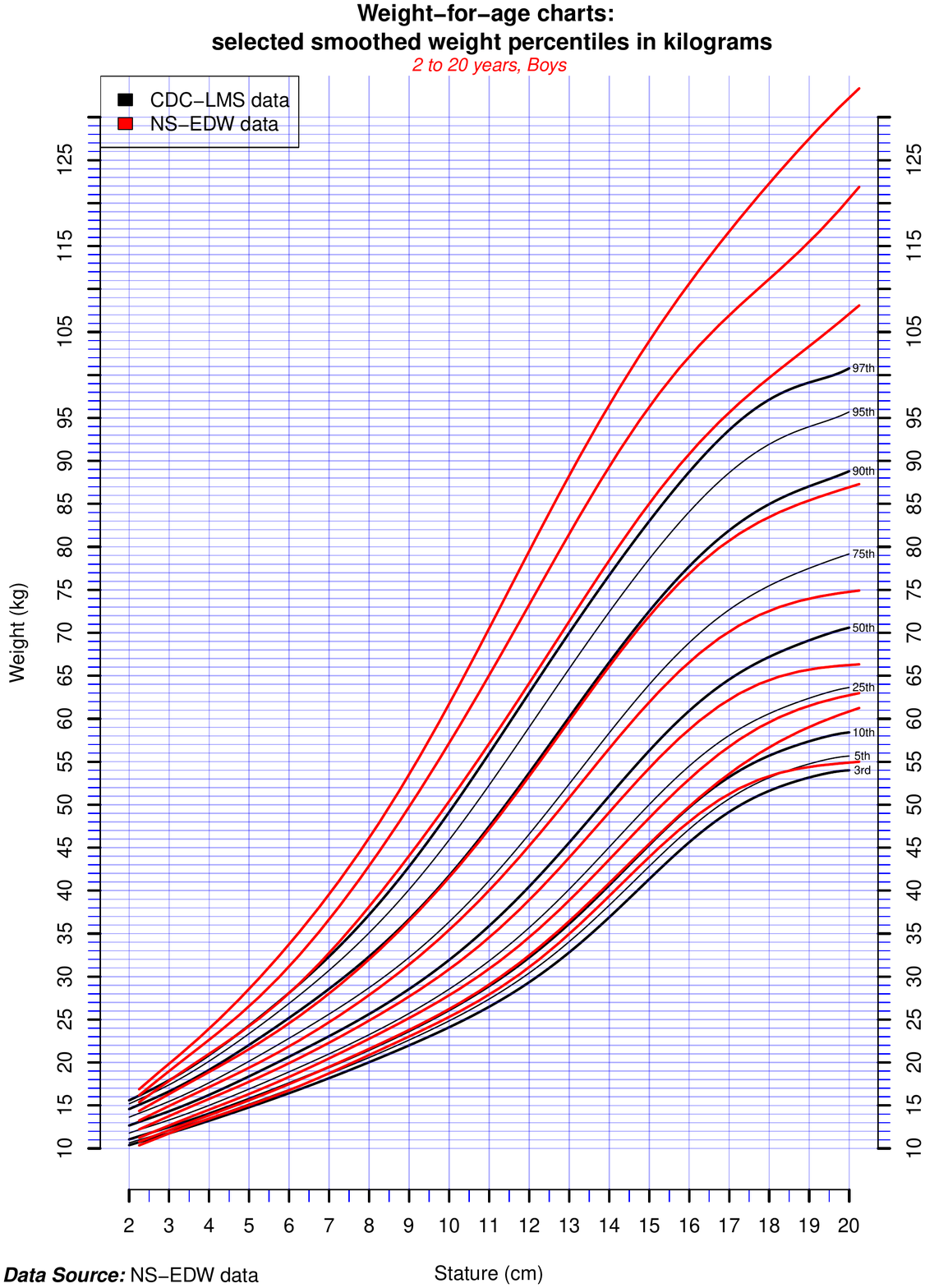}
}
\hfil
\subfloat[NHANES]{\includegraphics[width=3.5in]{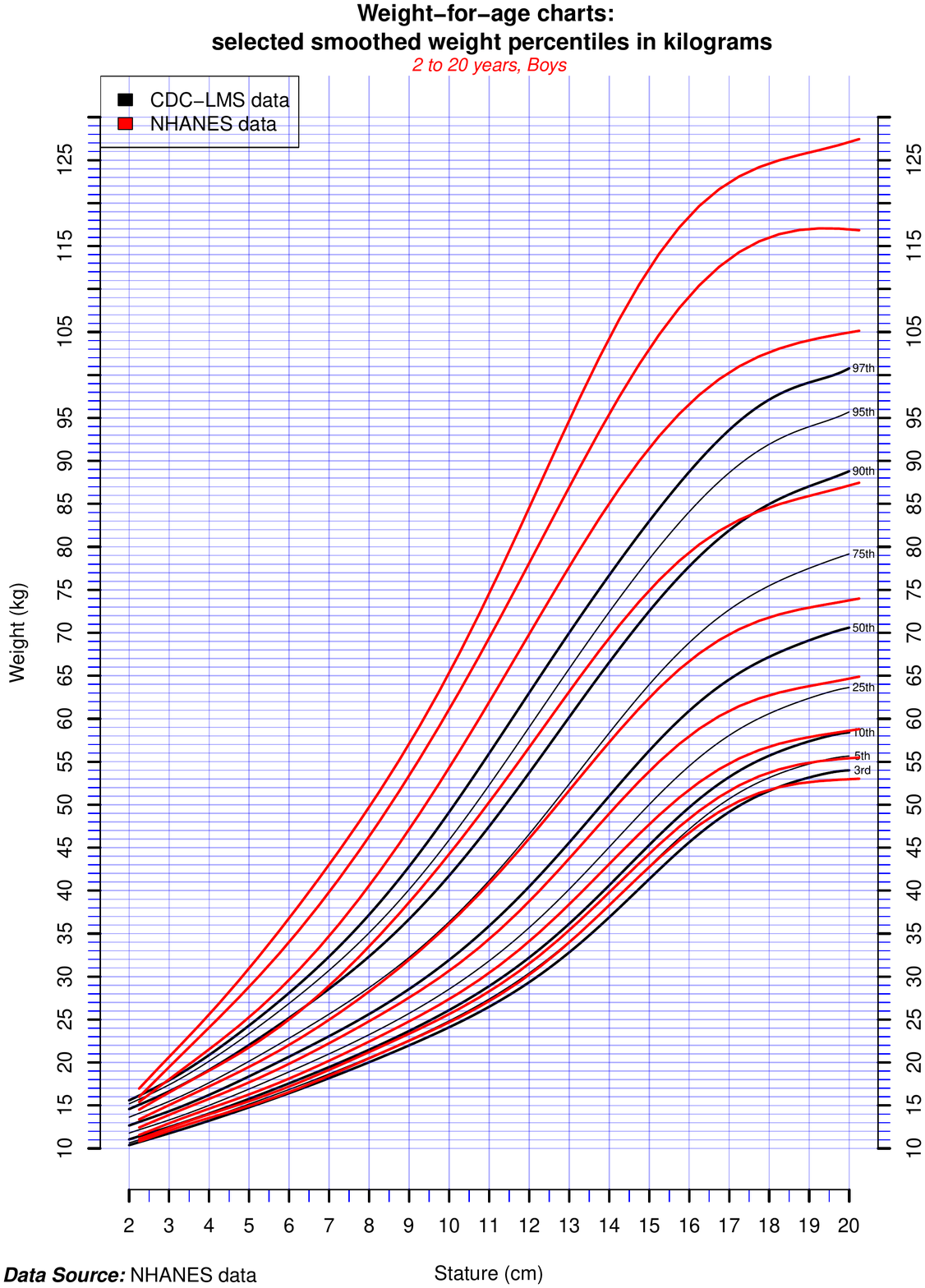}
}}
\caption{Boys Weight-for-age: 2-20 years}
\label{fig_weight}
\end{figure*}

\begin{figure*}[!t]
\centerline{\subfloat[NS-EDW]{\includegraphics[width=3.5in]{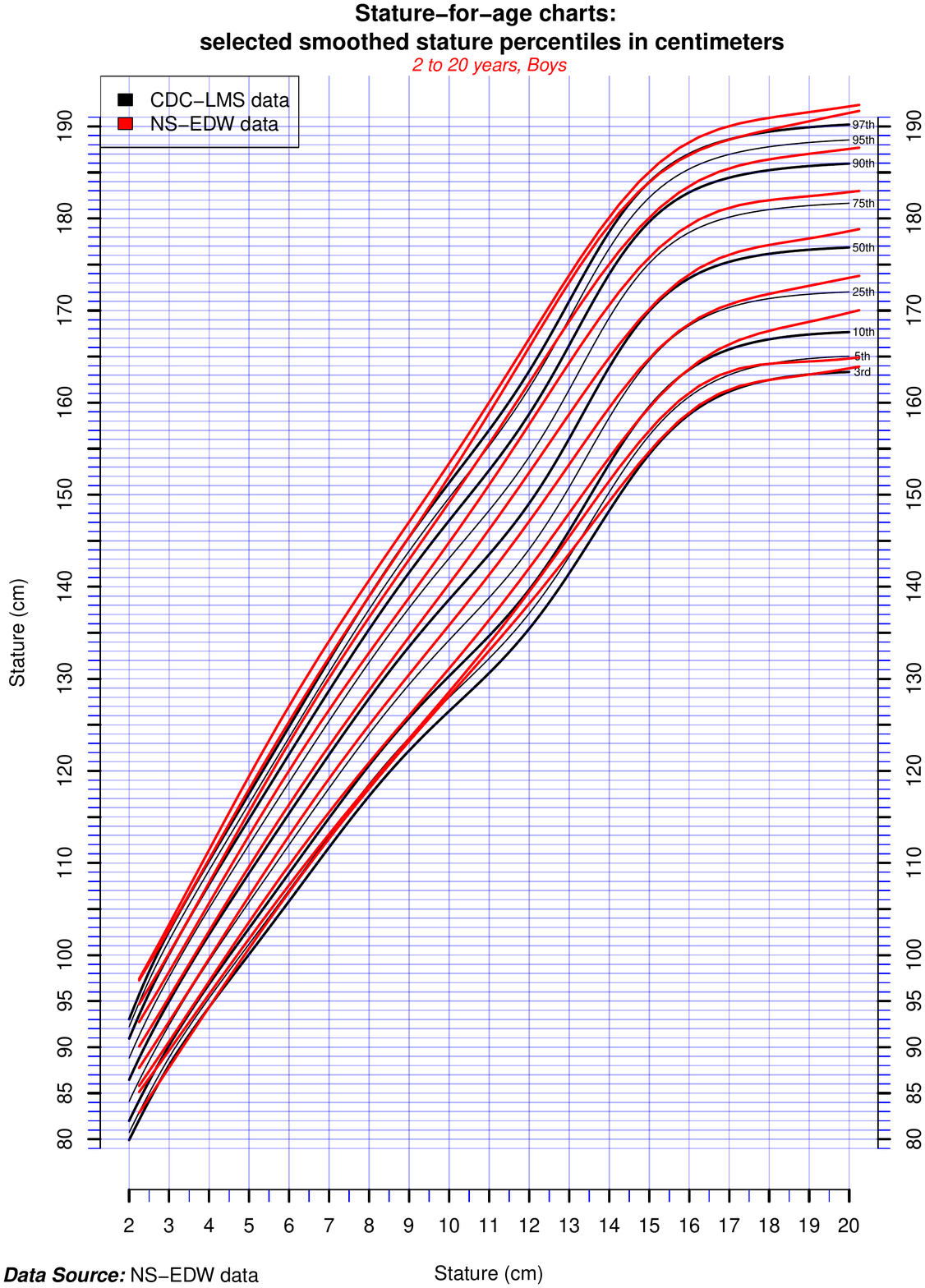}
}
\hfil
\subfloat[NHANES]{\includegraphics[width=3.5in]{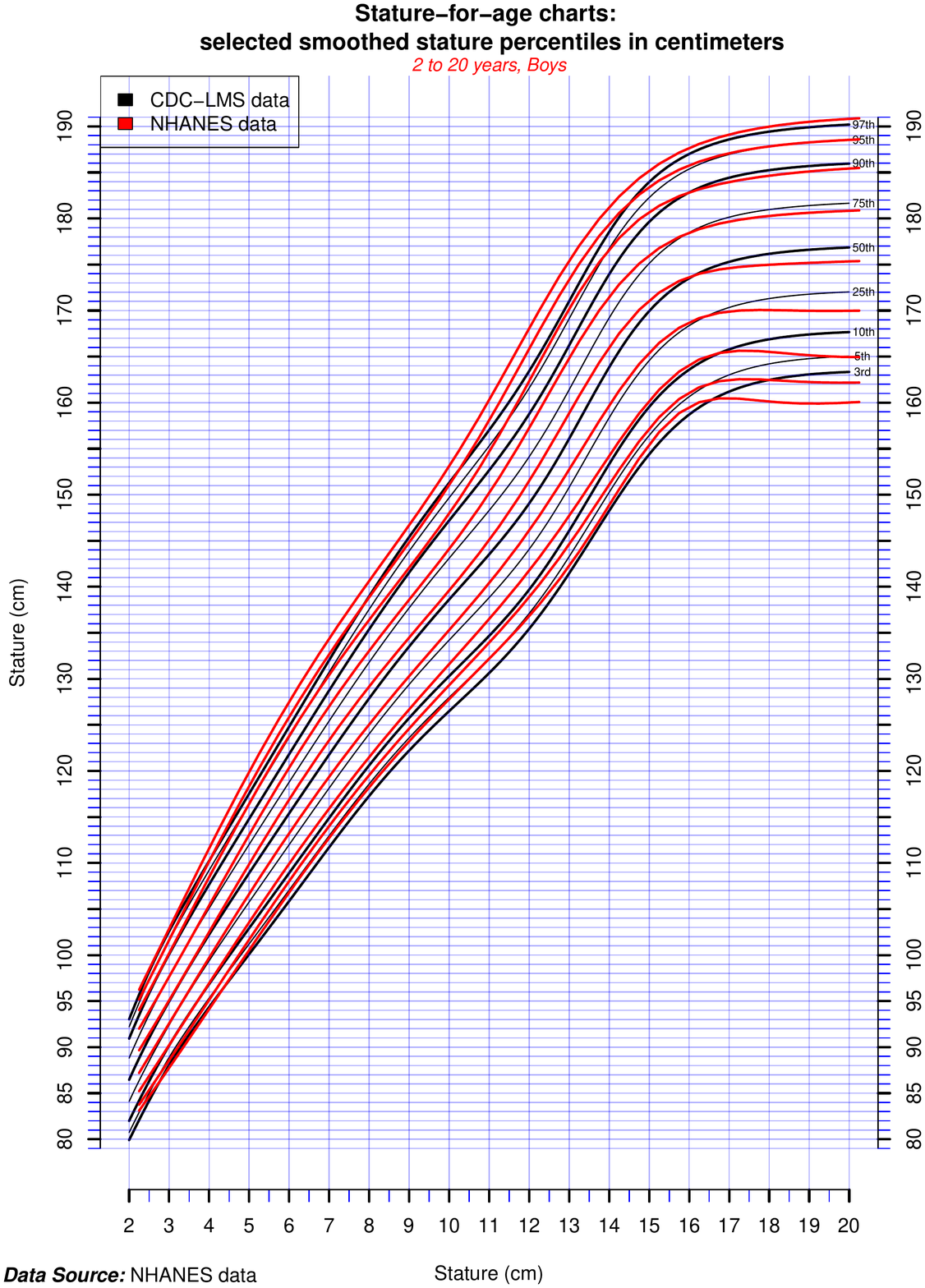}
}}
\caption{Boys Stature-for-age: 2-20 years}
\label{fig_stature}
\end{figure*}

\begin{figure*}[!t]
\centerline{\subfloat[NS-EDW]{\includegraphics[width=3.5in]{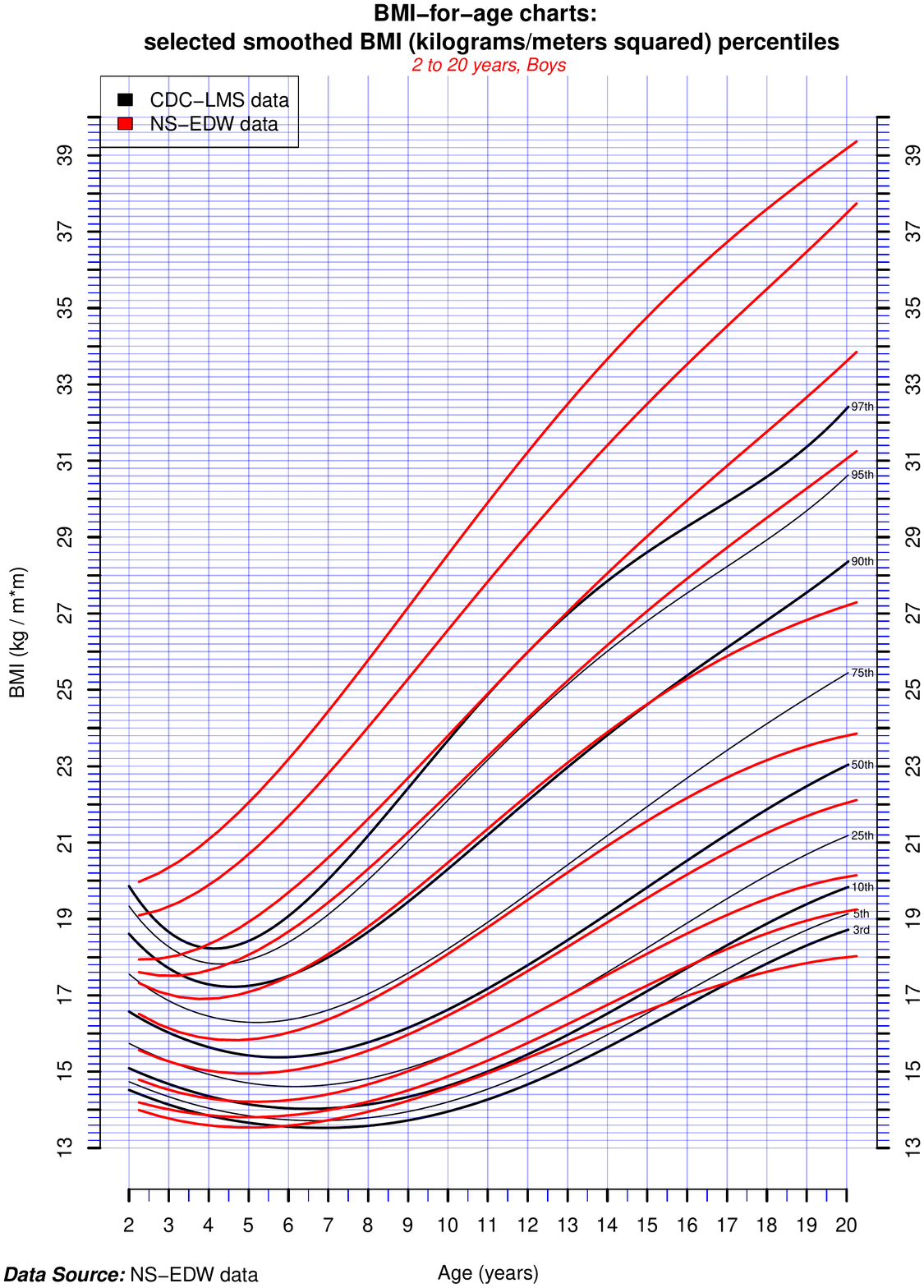}
}
\hfil
\subfloat[NHANES]{\includegraphics[width=3.5in]{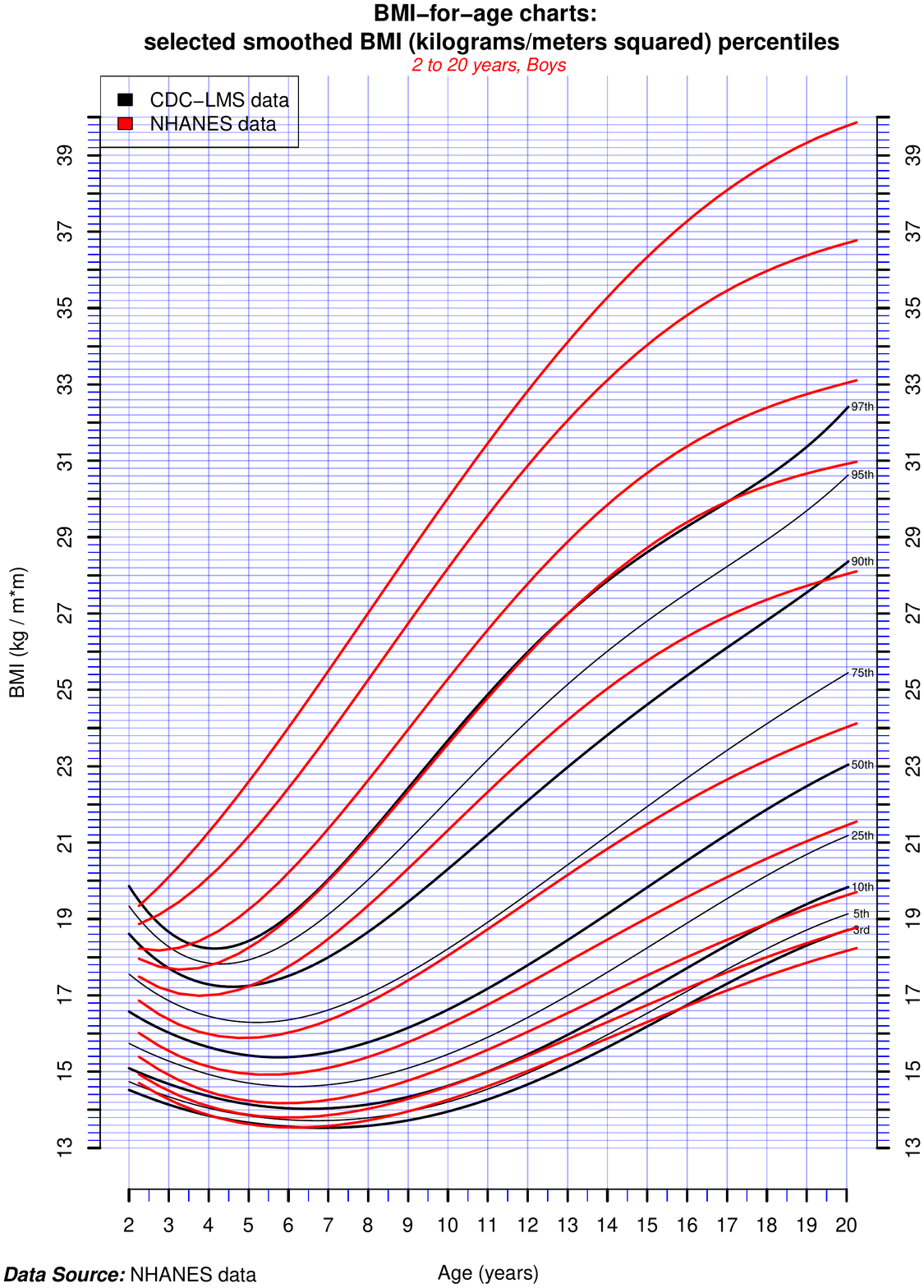}
}}
\caption{Boys BMI-for-age: 2-20 years}
\label{fig_bmi}
\end{figure*}

\begin{figure*}[!t]
\centerline{\subfloat[NS-EDW]{\includegraphics[width=3.5in]{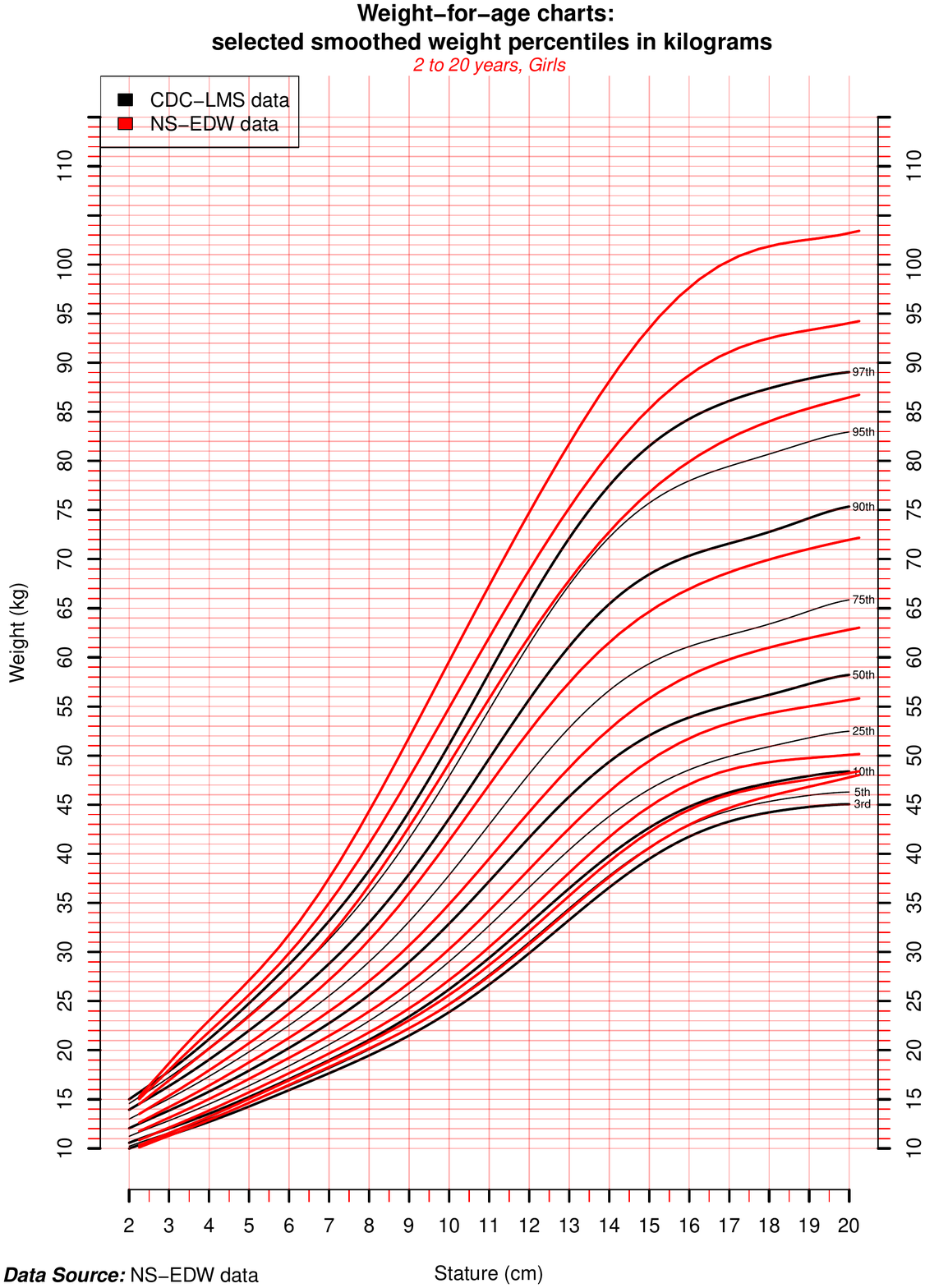}
}
\hfil
\subfloat[NHANES]{\includegraphics[width=3.5in]{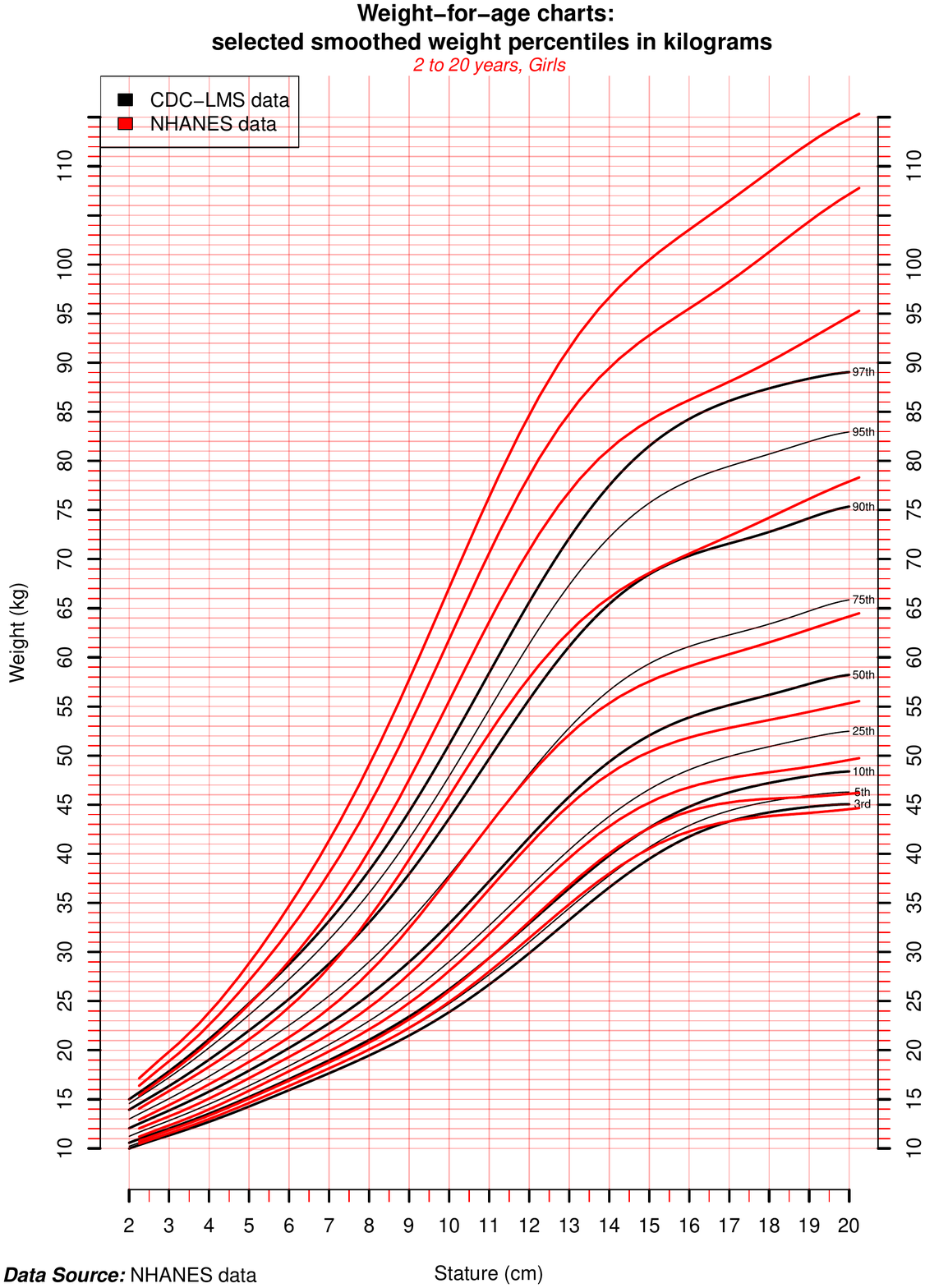}
}}
\caption{Girls Weight-for-age: 2-20 years}
\label{fig_weight_g}
\end{figure*}

\begin{figure*}[p]
\centerline{\subfloat[NS-EDW]{\includegraphics[width=3.5in]{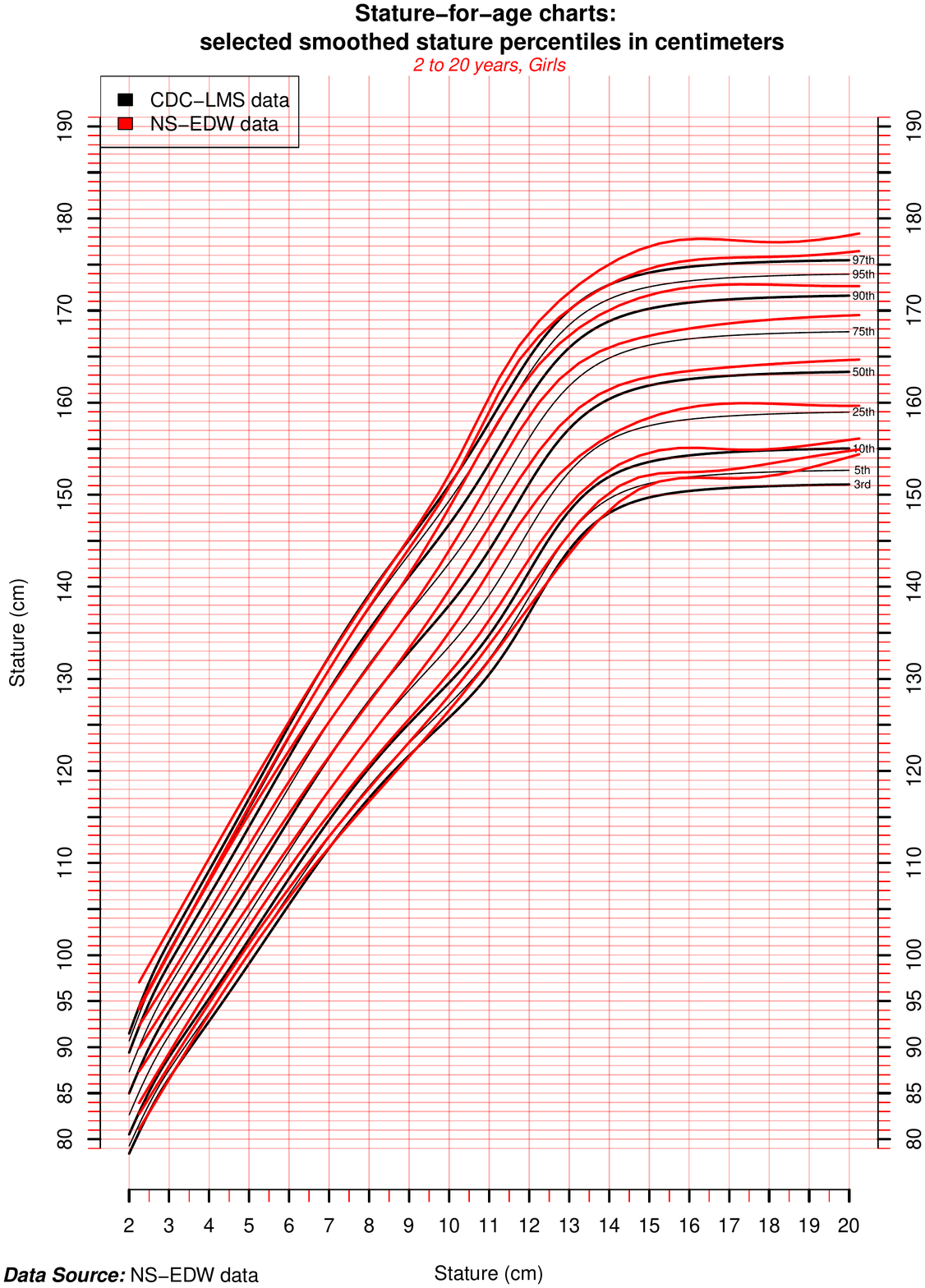}
}
\hfil
\subfloat[NHANES]{\includegraphics[width=3.5in]{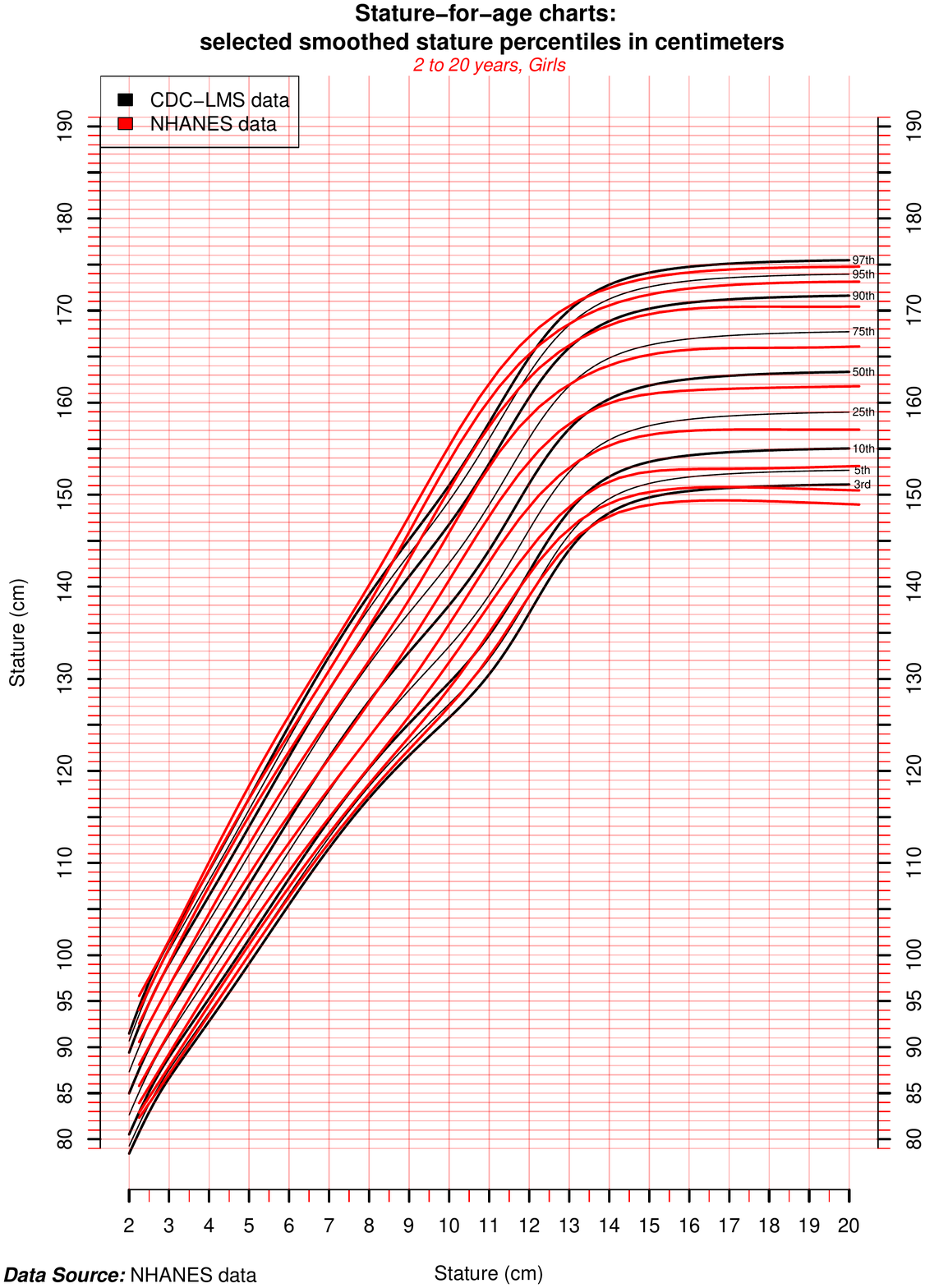}
}}
\caption{Girls Stature-for-age: 2-20 years}
\label{fig_stature_g}
\end{figure*}

\begin{figure*}[!t]
\centerline{\subfloat[NS-EDW]{\includegraphics[width=3.5in]{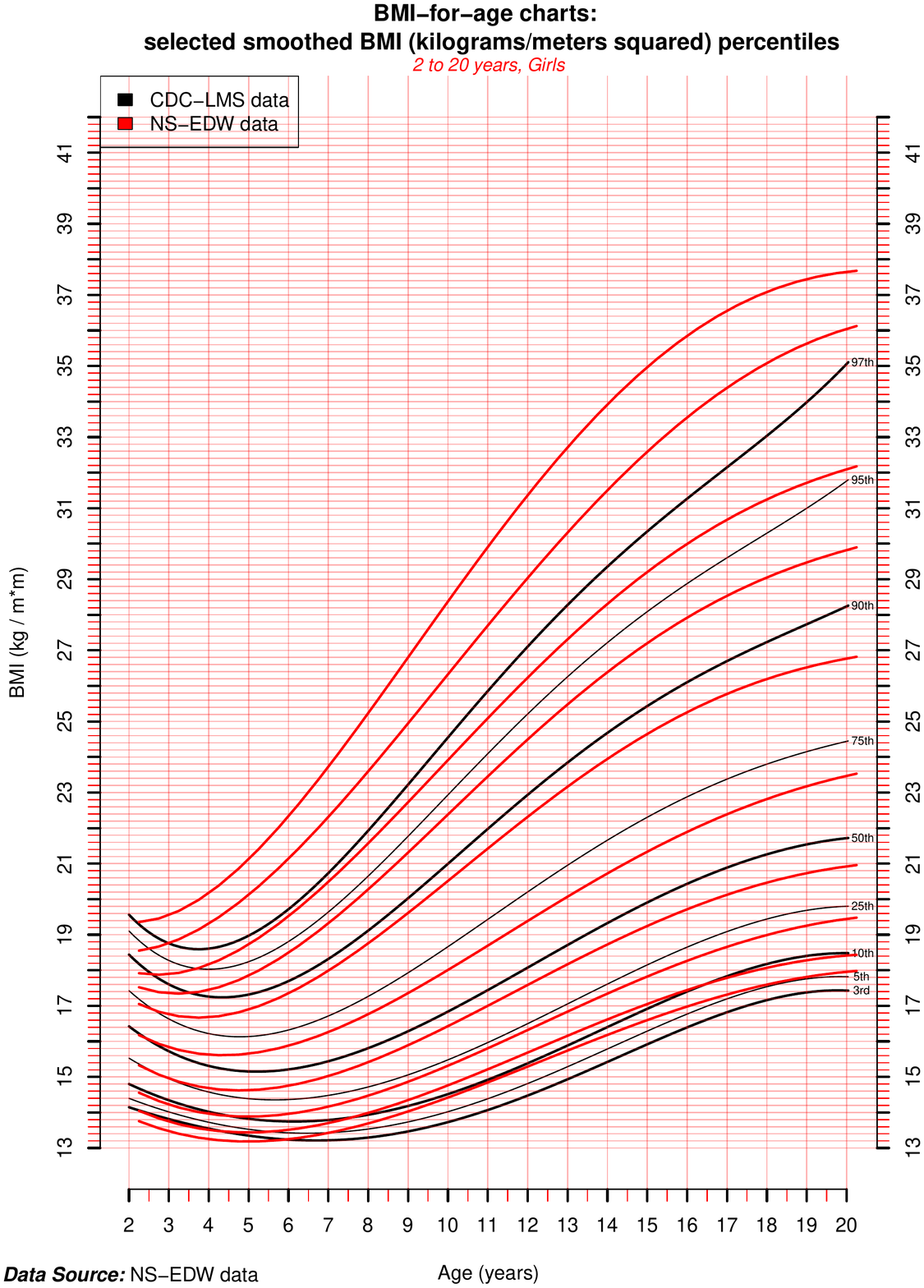}
}
\hfil
\subfloat[NHANES]{\includegraphics[width=3.5in]{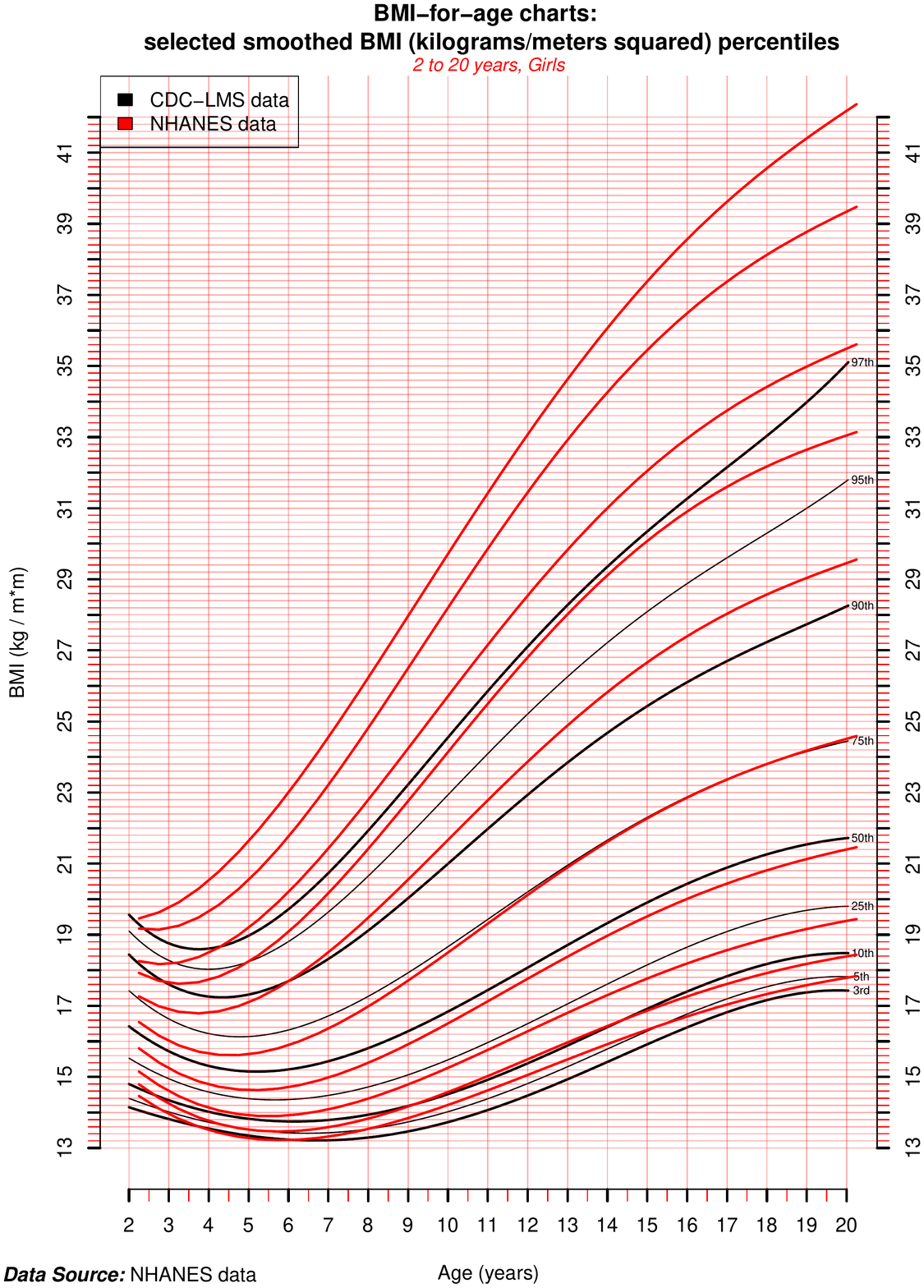}
}}
\caption{Girls BMI-for-age: 2-20 years}
\label{fig_bmi_g}
\end{figure*}

Now that we concluded the success in reproducing the curve smoothing stages described in \cite{CDCpaper02}, we followed the methods and generated the smoothed percentile curves for NS-EDW data and NHANES data and compared with the published CDC curves. The twelve growth charts of weight-for-age, stature-for-age and BMI-for-age of boys and girls are shown in Figure~\ref{fig_weight} through Figure~\ref{fig_bmi_g}. The curves in black were generated by the published LMS value from the CDC website (see \cite{CDCdatawebsite}) and the curves in red are generated from NS-EDW/NHANES data set correspondingly [See page 8-10 for Figure~\ref{fig_weight} - Figure~\ref{fig_bmi} and page 11-13 for Figure~\ref{fig_weight_g} - Figure~\ref{fig_bmi_g}].

\begin{figure}[!t]
\centering
\includegraphics[width=4.5in]{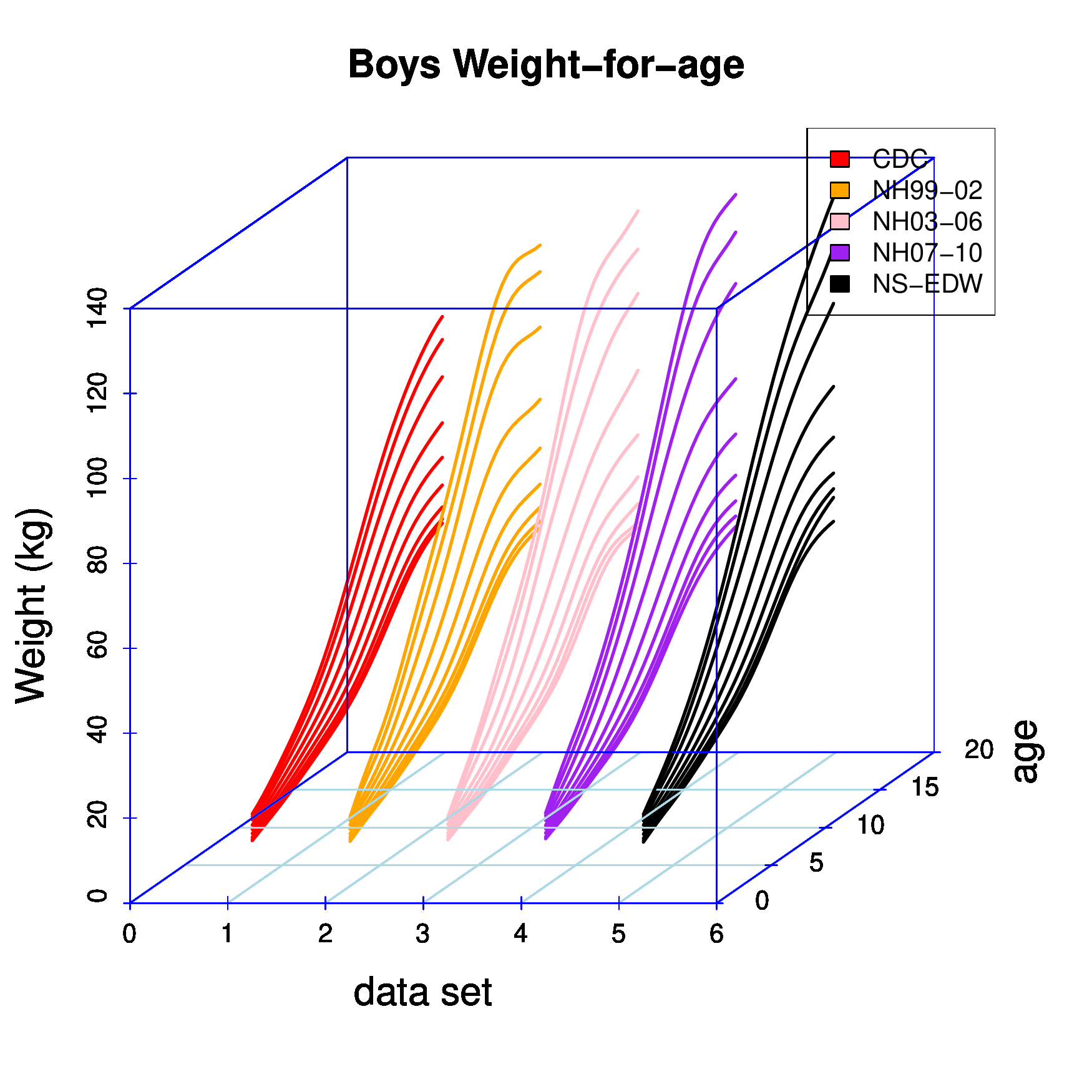}
\caption{Boys Weight-for-age Comparison}
\label{fig_3dcomparison}
\end{figure}

\begin{figure}[!t]
\centering
\includegraphics[width=4.5in]{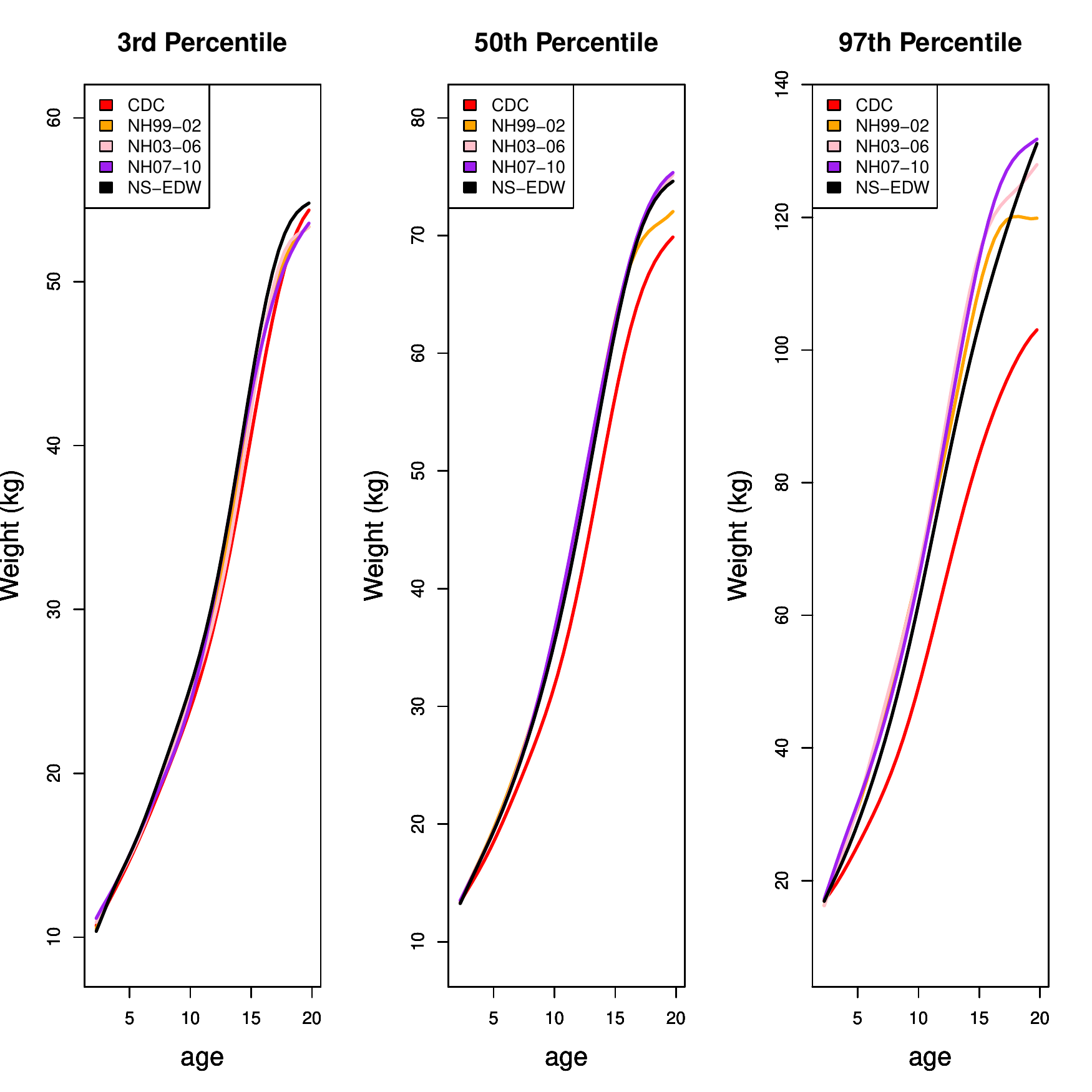}
\caption{Selected Weight-for-age Percentiles}
\label{fig_comparison}
\end{figure}

As an example illustrating the fattening trends by year of American children, the boys weigh-for-age percentile curves for the CDC data, the three non-overlapping four-yearly NHANES data and the NS-EDW data (labeled as 1 to 5 in chart) are plotted in Figure~\ref{fig_3dcomparison}. The curves showed an ascending trend from data set 1 to 5. To better illustrate the difference more clearly, three selected representative percentiles (3rd, 50th and 97th) are compared and shown in Figure~\ref{fig_comparison}. [See page 14 for Figure~\ref{fig_3dcomparison} - Figure~\ref{fig_comparison}].

\section{Discussions}
As summarized in the Figure~\ref{fig_timeline}, though the CDC growth charts were released in the year 2000, the data used to generate these is considerably older. For younger children they date from the 1960s and 1970s up to 1994. For older children they are only from the 1960s and 1970s. Also the CDC charts are not truly 'growth' data because they are cross-sectional, not longitudinal. The NHANES data collected from 1999 to 2010 covers a wider range of children and is much more representative of the current generation. The NS-EDW data provided data collected from 2006 to mid 2012 from routine office visits. This data is longitudinal and can be used for studies related to early adiposity rebound in children.
\subsection{Stature-for-age}
Comparisons of boys and girls stature-for-age charts from the CDC data, NS-EDW data and NHANES data show that these curves do not differ much, i.e. the stature growth pattern of the American children has been fairly consistent over the past 50 years.

\subsection{Weight-for-age}
The weight-for-age of both NS-EDW and NHANES differ substantially from the CDC data. The curves generated from both data sets are shifted upwards, indicating a progressive increase in American children's weight for both girls and boys through these years. The upward shifts also become more and more significant with age. The accelerating trends for boys weight of the NS-EDW and NHANES data are similar while the girls weight of the NHANES data increased even more significant than the NS-EDW girls.

\subsection{BMI-for-age}
The BMI-for-age of both NS-EDW and NHANES also differ from the CDC data. As illustrated before, the children's stature growth pattern does not change much while the weight increased substantially. Since BMI is directly proportional to weight, the increase of BMI is natural.

However, the shape of the BMI-for-age curves of both NS-EDW and NHANES also changed. BMI in childhood typically increases through the first year of life and then drops off before reaching a minimum later in childhood and then increasing again, as shown in the CDC BMI-for-age charts. This phenomenon is known as adiposity rebound (AR) \cite{Rolland87}.

However as shown in the NHANES and NS-EDW curves, there is no obvious adiposity rebound phenomenon for the 95th and 97th percentiles of age 2-20 years. Unlike the CDC data, the nadir of those percentiles are not very obvious. This might partly due to the insufficiency of accurate sample for children age 1 to 2 years. However, adiposity rebound points for other percentiles also left shifted, consistent with the observations for 95th and 97th percentiles (whose AR could be regarded as occurring even before age 2 years).

Previous studies have shown the mean age of adiposity rebound for boys and girls to be 6.6 years, and 6.0 years respectively \cite{Williams05,Janicke09}. A substantial body of evidence indicates that children in whom adiposity rebound occurs early (before age 5.5 for boys and 5.0 for girls) are at much higher risk for obesity and its consequences in later childhood and young adulthood compared with children for whom adiposity rebound occurs later \cite{Janicke09}. For children with early adiposity rebound compared with average adiposity rebound, the relative risk at age 26 of being overweight is 2.70 and the relative risk of obesity is 5.91 \cite{Janicke09}. A recent study of growth patterns among massively obese French adolescents revealed an average age of adiposity rebound of just 2 years \cite{Waters11}, which is similar to our findings in respect with the 95th and 97th percentiles. As early adiposity rebound is a risk factor for future obesity, these findings indicate the increasing risk for the population wide obesity.

\section{Conclusions}
In this paper, we created selected empirical percentiles of growth charts for data from the NorthShore Enterprise Data Warehouse (NS-EDW) and from the National Health and Nutrition Examination Survey (NHANES) by closely following the methods used to generate the 2000 CDC growth curves. Weight-for-age, stature-for-age and BMI-for-age for boys and girls of age 2-20 were generated and compared with the 2000 CDC growth curves. The results show that the growth pattern of stature-for-age is similar to the CDC data. The weight-for-age and BMI-for-age percentile curves differ substantially from the CDC percentile charts. The shape of the BMI-for-age curves are different from the CDC curves. We conclude that in the United States, the weight and BMI values are increasing, there is a progressive fattening of American children. The growth charts generated in the early years as standards for measuring growth might no longer be applicable to today's population of American children.

In future, we plan to closely study the characteristic patterns of BMI-for-age curves from the longitudinal data available through NS-EDW and discover early predictive factors of obesity, such as adiposity rebound. To date, early adiposity rebound has been identified exclusively in research studies in which collection of BMI data to identify growth patterns is systematically planned. However, no studies have made use of clinical data collected from typical, ambulatory practices. Identifying early adiposity rebound from clinical practice data may be useful, which encourage clinicians to monitor BMI more carefully in children with early rebound and to provide additional counseling.

From the Electronic Medical Records in NS-EDW, fully anonymized versions of routinely measured children's weight and stature are available for research. The longitudinal nature of this data can be used to investigate individual children's growth curves and their relationships with population growth characteristics presented in this paper. Through these efforts, we plan to identify the presence of early adiposity rebound, to estimate the timing of early adiposity rebound, and use these to prevent childhood obesity.

\section*{Acknowledgment}
We thank Cynthia Ogden for helping access the CDC data, John Komlos for helping access the NHANES data, Chad Konchack and Justin Lakeman for extracting the NS-EDW data. We have benefited from conversations with and comments from Katherine Flgal, Tim Sanborn, Kibaek Kim, Joyce Ho, Sanjay Mehrotra, Tony Solomonides, Yuan Ji, Nigel Parsad, and Jonathan Silverstein.


\begin{thebibliography}{1}
\bibitem{Ogden10}
C.~L. Ogden, M.~D. Carroll, L.~R. Curtin, M.~L.Lamb, K.~M. Flegal, "Prevalence of high body mass index in US children and adolescents 2007-2008," JAMA 2010; 303(3): 242-9.
\bibitem{CDCpaper02}
R.~J. Kuczmarski, C.~L. Ogden, S.~S Guo, et al., "2000 CDC growth charts for the United States: Methods and development," National Center for Health Statistics. Vital Health Stat 2002;11(246).
\bibitem{CDCwebsite}
$http://www.cdc.gov/growthcharts/$
\bibitem{Owen78}
G.~M. Owen, "The new National Center for Health Statistics growth charts," South Med J 1978; 71:296¨C7.
\bibitem{WHO}
WHO Multicentre Growth Reference Study Group. WHO child growth standards based on length/height, weight and age. Acta Paediatr 2006; 76-85.
\bibitem{Cole98}
T.~J. Cole, J.~V. Freeman, M.~A. Preece, "British 1990 growth reference
centiles forweight, height, body mass index and head circumference
fitted by maximum penalized likelihood," Stat Med 1998;17:407-29.
\bibitem{Cole99}
T.~J. Cole, M.~J. Roede, "Centiles of body mass index for Dutch children aged 0-20 years in 1980¡ªa baseline to assess recent trends in
obesity," Ann Hum Biol 1999;26:303-8.
\bibitem{Cole88}
T.~J. Cole, "Fitting smoothed centile curves to reference data," J R Stat Soc 1988; 151:385¨C418.
\bibitem{Cole92}
T.~J. Cole, P.~J. Green,  "Smoothing reference centile curves: The LMS
method and penalized likelihood," Stat Med 1992;11:1305¨C19.
\bibitem{Flegal13}
K.~M. Flegal, T.~J. Cole, "Construction of LMS parameters for the Centers for disease Control and Prevention 2000 growth charts," Hational health statitics reports; no 63. Hyattscille, MD: National Center for Health Statistics. 2013.
\bibitem{ChenSAS}
C.~Chen, Growth Charts of Body Mass Index (BMI) with Quantile Regression. SAS Institure Inc. Cary, NC, USA.
\bibitem{Mei02}
Z.~Mei, L.~M. Crummer-Strawn, A.~Pietrobelli, A.~Coulding,  M.~I. Goran, W.~H. Dietz, "Validity of body mass index compared with other body-composition screening indices for the assessment of body fatness in children and adolescents," AJCN 2002;7597-985.
\bibitem{Rolland84}
M.~F. Rolland-Cachera, M.~Deheeger,  F.~Belisle, et al., "Adiposity rebound in children: a simple indicator for predicting obesity," AJCN 1984;39:129-35.
\bibitem{Rolland87}
M.~F. Rolland-Cachera, M.~Deheeger, P.~Avons,  M.~Cuilloud-Bataille, E.~Patois, M.~Sempe, "Tracking adiposity patterns from 1 month to adulthood," Ann Hum Biol 1987;14:219-22.
\bibitem{Taylor05}
R.~W. Taylor, A.~M. Grant,  A.~Goulding, S.~M. Williams, "Early adiposity rebound: review of papers linking this to subsequent obesity in children and adults," Curr Opin Clin Nure Metab Care 2005; 8: 607-12.
\bibitem{Flegal09}
K.~M. Flegal, R.~Wei, C.~L. Ogden, D.~S. Freedman, C.~L. Johnson, L.~R. Curtin, "Characterizing extreme values of body mass index-for-age by using the 2000 Centers for Disease Control and Prevention growth charts," Am J Clin Nutr 2009; 90(5):1314-20.
\bibitem{Tim07}
T.~J. Cole, K.~M. Flegal,  D.~Nicholls,  A.~A. Jackson, "Body mass index cut offs to define thinness in children and adolescents: international survey," BMJ 2007;335:194.
\bibitem{Rao13}
G.~Rao, "A new tool needed for identifying and characterizing obesity," Curr Cardiovasc Risk Rep. 2013.
\bibitem{CDCnhaneswebsite}
$http://www.cdc.gov/nchs/nhanes/about\_nhanes.htm$
\bibitem{Rwebsite}
$http://statistic-on-air.blogspot.com/2011/09/implementation-of-cdc-growth-charts-in.html$
\bibitem{Barlow98}
S.~E. Barlow, W.~H. Dietz, "Obesity evaluation and treatment: Expert committee recommendations," Pediatrics 1998;102(3):e29.  URL: $http://www.pediatrics.org/cgi/content/full/102/3/e29$
\bibitem{Himes94}
J.~H. Himes, W.~H. Dietz, "Guidelines for overweight in adolescent preventive services: Recommendations from an expert committee," Am J Clin Nutr 1994;59:307-16.
\bibitem{Cleveland79}
W.~S. Cleveland  "Robust locally weighted regression and smoothing scatterplots," JASA 79:829-36.1979.
\bibitem{CDCdatawebsite}
$http://www.cdc.gov/growthcharts/percentile\_data\_files.htm$
\bibitem{Williams05}
S.~M. Williams, "Weight and height growth rate and the timing of adiposity rebound," Obes Res. 2005:13:1123-1130.
\bibitem{Janicke09}
D.~M. Janicke, B.~J. Sallinen,  M.~G. Perri, L.~D. Lutes, J.~H. Silverstein, B. ~Brumback, "Comparison of program costs for parent-only and family-based interventions for pediatric obesity in medically underserved settings," J Rural Health 2009; 79(7):319-25.
\bibitem{Waters11}
E.~Waters, A.~de Silva-Sanigorski, B.~J. Hall, T.~Brown, K.~J. Campbell, Y.~Gao, R.~Armstrong, L.~Prosser, C.~D. Summerbell,  "Interventions for preventing obesity in children," Cochrane Database Syst Rev 2011;12: CD001871.
\end{thebibliography}
\end{document}